\documentclass{kapproc}  
\usepackage{procps}
\usepackage[dvips]{graphicx}
\upperandlowercase

\nochapequationnumber

\let\footnote\savefootnote

\let\footnotetext\savefootnotetext

\normallatexbib

\begin{document} 
%
%
%
%
\articletitle{Signals \ of \ Deconfinement \ Transition \ 
 in \ Nucleus-Nucleus \ Collisions
 }
\author{Mark I. Gorenstein}
\affil{Bogolyubov Institute for Theoretical Physics, Kiev, Ukraine} 
\email{mark@mgor.kiev.ua}
 
\begin{abstract}
We discuss the energy dependence of hadron production  in relativistic
nucleus-nucleus collisions. 
Several `anomalies' in the energy dependence have been predicted as
signals of the deconfinement phase transition and
three of these signals are observed at the CERN SPS indicating that the
onset of deconfinement in Pb+Pb collisions is located at about 
30~A$\cdot$GeV. 
\end{abstract}


\section{Introduction}\label{intro}
\noindent
The data on heavy nucleus-nucleus (A+A) collisions suggested   
that there is a significant
change in the energy dependence of the pion and strangeness
yields which is located between the top AGS (11~A$\cdot$GeV) and SPS
(158~A$\cdot$GeV) energies.
Based on the statistical approach 
it was speculated that the change is related to the
onset of deconfinement at the early stage of A+A
collisions, and a simplified quantitative model was developed,
the Statistical Model of the Early Stage (SMES) \cite{GaGo}.
The SMES predicted a sharp maximum in the multiplicity ratio of strange
hadrons to pions at the begining of the transition region.
This prediction triggered a new experimental programme at the SPS -- the
energy scan programme -- in which the NA49 experiment recorded central  
Pb+Pb collisions at several energies:
the results from the run at 40, 80 and 158~A$\cdot$GeV are 
published \cite{NA49a},  the results from the 30~A$\cdot$GeV
run were shown
for the 
first time at the conference SQM03 \cite{NA49b},
the data at 20~A$\cdot$GeV are still being analysed.
The energy scan program at the CERN SPS results an observation of several
`anomalies' in the energy dependence of hadron production in the same
domain of the collision energy. 
These `anomalies' are interpreted
as signals of the deconfinement phase transition 
in A+A collisions.
In this report we review
the physical arguments which lead us to the proposed signals
as well as discuss their experimental status.

\section{Signals of Deconfinement: Model Predictions}
\noindent
Originally, two signals of the deconfinement phase
transition were proposed within SMES \cite{GaGo}:
the energy dependence of the mean pion and
strangeness multiplicities.
Recently, two new signals were suggested within
SMES:
the energy dependence of the inverse slope of the
transverse mass spectrum of kaons
\cite{GoGaBu} and the energy dependence of properly filtered
multiplicity fluctuations \cite{GaGoMo,GoGaZo}.

An exact nature of the deconfinement phase transition is still debated.
On the other hand, it is rather well established in the lattice QCD at
zero baryonic chemical potential that
very strong changes of the energy density $\varepsilon$ take place
in a narrow temperature interval $\Delta T = 5\div 10$~MeV.
Within this temperature interval the energy density
changes by  about an order of magnitude, whereas the pressure remains
approximately  unchanged. One may refer to this temperature interval as a
`generalised mixed phase'. 
 
\vspace{0.2cm}
\noindent
{\bf The Pion Multiplicity.}~~~
The majority of all particles produced
in high energy interactions are pions.
Thus, pions carry basic information on the entropy
created in the collisions.
On the other hand, the entropy production should
depend on the form of matter present at the early stage
of collisions.  
A deconfined matter is expected to lead to the 
final state with higher entropy than that created by
a confined matter.
Consequently, it is natural to expect that the onset of
creation of a deconfined matter should be signalled by
an enhancement of the pion production.
Clearly, a trivial dependence of the pion multiplicity on the
size of colliding nuclei should be removed and thus
a relevant observable is the ratio of the mean pion multiplicity
$\langle \pi \rangle$ to the  mean number of wounded nucleons
$\langle N_W \rangle$.  
The simple intuitive argumentation can be further quantified
within SMES assuming  the generalised Fermi--Landau initial
conditions: the initial volume is Lorenz-contracted, $V\propto
(\sqrt{s})^{-1}$ ($\sqrt{s}$ is the c.m.s. energy
of the nucleon pair), the initial energy density is given by
$\varepsilon \propto gT^4\propto (\sqrt{s} - 2m_N)\cdot \sqrt{s}$ ($T$ is
the initial
temperature and $g$ is an effective
number of internal degrees of freedom  at the early stage, $m_N$ is the
nucleon mass).
The pion multiplicity is proportional to the initial entropy, and
the $\langle \pi \rangle/\langle N_W \rangle$
ratio can be thus calculated outside the transition region as
\begin{equation}\label{npi}
\langle \pi \rangle/\langle N_W \rangle~\propto V~ g ~T^3 ~
\propto g^{1/4}  (\sqrt{s} - m_N)^{3/4} (\sqrt{s})^{-1/4}  ~\equiv
~g^{1/4}~F~.
\end{equation}
Therefore, the $\langle \pi \rangle/\langle N_W \rangle$
ratio increases linearly with $F$  outside the transition region, and  the
slope parameter is proportional to
$g^{1/4}$ \cite{Ga}.
In the transition region, a steepening of the
pion energy dependence is predicted, because of activation of 
partonic degrees of freedom, i.e. an effective
number of internal degrees of freedom in the quark gluon plasma 
(QGP) is
larger than in the
hadron gas (HG): $g_{QGP}>g_{HG}$.

\vspace{0.2cm}
\noindent
{\bf The Strangeness to Pion Ratio.}~~~~ 
The energy dependence of the strangeness to entropy
ratio is a crucial  signal of the deconfinement
due to its weak dependence on the
assumed initial conditions.
Within SMES 
at low collision energies, when the confined matter is produced, the
strangeness to entropy ratio steeply increases with collision energy.
Due to a low temperature at the early stage ($T < T_C\cong 170$~MeV)
and the high mass of the carriers of strangeness ($m_S \cong
500 $~MeV, the kaon mass) the total strangeness is 
$\propto
\exp(-m_S/T)$. On the other hand, the total entropy is approximately
$\propto T^3$. Therefore, the strangeness to pion ratio is $\propto
\exp(-m_S/T)\cdot T^{-3}$ in the HG and strongly increases with the
collision energy. When the transition to a
deconfined matter is crossed ($T > T_C$),
the mass of the strangeness carriers is
significantly reduced ($m_s =130 \div 170$ MeV, the strange quark mass).
Due to the low mass ($m_s < T$) the strangeness yield becomes
(approximately)
proportional to the entropy (both are proportional to $T^3$), and the
strangeness
to entropy (or pion) ratio becomes independent of energy in the QGP.
This leads to a ``jump'' in the energy
dependence from the larger value for confined matter at $T_C$ to the
value for deconfined matter.
Thus, within the SMES, the non-monotonic energy
dependence of the strangeness to entropy ratio is followed by a
saturation at the deconfined value
which is a direct consequence of the onset of deconfinement taking place
at about 30~$A$GeV \cite{GaGo}.

\vspace{0.2cm}
\noindent
{\bf The Inverse Slope of Transverse Mass Spectra.}~~~~
 We discuss another well known
observable, which may be sensitive to the onset of deconfinement,
the transverse momentum, $p_T$, spectra of produced hadrons.
It was suggested by Van Hove
\cite{van-hove} more than 20 years ago to identify the deconfinement phase
transition in high energy proton-antiproton interactions with
a plateau-like structure of the average transverse momentum as
a function of the hadron multiplicity\footnote{In the original 
suggestion \cite{van-hove} the correlation between average
transverse momentum and hadron
multiplicity was discussed for $p+\bar{p}$ at
fixed collision energy. Today we have an advantage to use central A+A
collisions at different energies \cite{GoGaBu}.}.
According
to the general concepts of the hydrodynamical approach the hadron
multiplicity reflects the entropy, whereas
the transverse hadron activity reflects
the combined effects of temperature and collective transverse expansion.
The entropy is assumed to be created at the early stage of the collision
and is approximately constant during the hydrodynamic expansion. The
multiplicity  is proportional
to the entropy, $S=s \cdot V$, where $s$ is the entropy density and
$V$ is the effective volume occupied by particles.
During the hydrodynamic expansion, $s$
decreases and $V$ increases with $s \cdot V$ being
approximately constant. Large multiplicity  at high energies means a
large entropy density at the beginning of the expansion (and consequently
a larger volume at the end).
Large
value of $s$ at the early stage of the collisions
means  normally high  temperature $T$ at this stage.
This, in
turn, leads to an increase of the transverse hadron activity, a
flattening of the transverse momentum spectrum.
Therefore, with
increasing  collision energy
one expects to observe an increase of both the   
hadron multiplicity and the average transverse momentum per
hadron. However, 
presence of the deconfinement phase transition would change this
correlation. In the phase transition region, the initial entropy density
(and hence the final hadron multiplicity) increases with the collision
energy,
but temperature $T = T_C$ and pressure  $p = p_C$
remain constant.
The equation of state (EoS) presented in a form
$p(\varepsilon)/\varepsilon$ versus $\varepsilon$ shows a minimum (the
`softest
point' \cite{Hu:95}) at the boundary of the
{\it generalised mixed phase} and the QGP.
Consequently
the shape of
the $p_T$ spectrum
is approximately independent of the multiplicity or the collision
energy.   
The transverse
expansion effect may even decrease when  crossing
the transition region \cite{van-hove}.
Thus one expects an `anomaly' in
the energy dependence of the transverse
hadron activity: the average transverse momentum increases with
the collision
energy when the early stage matter is either in a pure confined or in
a pure
deconfined phase,
and it remains approximately constant when  the matter
is in a mixed phase \cite{GoGaBu,van-hove}.
A simplified picture with $T=T_{C}$ inside the mixed phase is
changed if the created early stage matter has
a non-zero baryonic density.
It was however demonstrated \cite{HS97} that the main
qualitative features ($T\cong const$, $p\cong const$,
and a minimum of the function
$p(\varepsilon)/\varepsilon$  versus  $\varepsilon$)
are present also in this case.
In the SMES model \cite{GaGo}, which  correctly predicted energy
dependence
of the pion and strangeness yields, the modification of the EoS due to
the deconfinement phase transition
is located between 30 and  about 160 $A\cdot$GeV.
Thus an anomaly in the energy dependence of 
the transverse hadron activity
may be expected in this energy range.

The energy
density at the early stage increases
with increasing collision energy.
At low and high energies, when a pure confined or
deconfined phase is produced, this leads to an
increase of the initial temperature and pressure.
This, in turn, results in an increase of the transverse
expansion of a matter and consequently a flattening
of the transverse mass spectra of final state hadrons.
The experimental data on the transverse 
mass spectra ($m_T= (m^2+p_T^2)^{1/2}$, $m$ is a particle mass) 
are usually parametrized by a simple exponential dependence:
\begin{equation}\label{T*} 
\frac{dN}{m_Tdm_T}~=~C~\exp\left(-~\frac{m_T}{T^*}\right)~,
\end{equation}
where the inverse slope parameter $T^{*}$
is sensitive to both the thermal and
collective motion in the transverse direction.
 In the parameterisation
(\ref{T*}), the shape of the $m_T$
spectrum is fully determined by
a single parameter, the
inverse slope $T^*$.
In particular,
the average transverse mass $\langle m_{T} \rangle$
can be expressed as:
\begin{equation}\label{mt}
\langle m_{T}\rangle ~=~ T^{*}~+~m~+~\frac{(T^{*})^{2}}{m~+~T^{*}}~.
\end{equation}
Hydrodynamical transverse flow with collective velocity  $v_T$ modifies
the Boltzmann $m_T$-spectrum of hadrons. At low transverse momenta, it 
leads to the
result ($T_{kin}$ is a kinetic freeze-out temperature):
\begin{equation}\label{T*1}
T^*_{low-p_T} ~=~T_{kin}~+~\frac{1}{2}
m~v_T^2~.
\end{equation}
A linear mass dependence (\ref{T*1}) of $T^*$ is supported by the data 
for hadron spectra at small $p_T$. However, for $p_T>>m$ the
hydrodynamical transverse flow leads
to the mass-independent blue-shifted `temperature':
\begin{equation}\label{T*2}
T^*_{high-p_T} ~=~T_{kin}~\cdot~ \sqrt{\frac{1+v_T}{1-v_T}}~. 
\end{equation}
A simple one parameter exponential fit (\ref{T*}) is quite accurate up
to $m_T -m \cong 1$~GeV for $K^{+}$ and $K^{-}$ mesons in A+A collisions
at
all energies. This means that $T^*_{low-p_T}\approx T^*_{high-p_T}$ for
kaons and
the energy dependence of the average
transverse
mass $\langle m_{T} \rangle$ (\ref{mt}) and the average transverse
momentum
$\langle p_{T} \rangle$ for kaons is
qualitatively the same as that for the parameter $T^{*}$.
Note 
that a simple exponential fit (\ref{T*}) neither  works for
light $\pi$-mesons, $T^{*}_{low-p_{T}} < T^{*}_{high-p_{T}}$, nor for
heavy (anti)protons
and (anti)lambdas, $T^{*}_{low-p_{T}} > T^{*}_{high-p_{T}}$. This means
that the
average transverse masses, $\langle m_{T}\rangle$, and their energy
dependence for these hadrons are not connected to the behavior of the
slope parameters in
the simple way described by Eq.~(\ref{mt}): one should separately consider
both $T^{*}_{low-p_{T}}$ and $T^{*}_{high-p_{T}}$ slopes 
(see Refs.~\cite{Sh,Go} for details).

\vspace{0.2cm}
\noindent
{\bf The Dynamical Event-by-Event Fluctuations.}~~~~
In thermodynamics, the energy $E$, volume $V$  and entropy $S$ are related
to each other
through the EoS. Thus, various values of the energy of
the initial equilibrium state lead to different, but uniquely determined,
initial entropies.
When the collision energy is fixed, the energy, which 
is used to hadron production
still fluctuates. These fluctuations of of the inelastic
energy are caused by the fluctuations in the dynamical
process which leads to the particle production.
They are called the dynamical energy fluctuations \cite{GaGoMo}.
Clearly the dynamical energy fluctuations lead to the dynamical
fluctuations of any macroscopic parameter of the matter, $X$,
like its entropy and strangeness content. 
The relation between the dynamical energy fluctuations and dynamical
fluctuations of the macroscopic parameter $X$ is given by the EoS.
 Consequently,
simultaneous measurements of the event-by-event fluctuations of both the
energy
and the parameter $X$ yield information on the EoS. Since EoS shows an
anomalous behavior in the
phase transition region, the anomaly should be visible in the ratio of
entropy to energy fluctuations \cite{GaGoMo}.

According to the first and the second principles of thermodynamics the
entropy change $\delta S$ is given as $T\delta S = \delta E + p \delta V$.
If we fix the collision geometry, choosing e.g. only a sample of
central A+A collisions, we can expect $ \delta V\cong 0$. 
Within SMES  the ratio of entropy to
energy fluctuations can be then easily calculated and presented as
a simple function of the $p/\varepsilon$
 ratio \cite{GaGoMo}:
\begin{equation}\label{R}
R_e ~\equiv ~\frac{(\delta S)^2/S^2}{(\delta E)^2/E^2}~=~
\left(1~+~\frac{p}{\varepsilon}\right)^{-2}~.
\end{equation}
Thus
it is easy to predict a qualitative dependence of the $R_e$
ratio on the collision energy.
 Within the model, the confined matter, which
is modelled as an ideal gas, is created at the collision early stage
below the energy of 30 A$\cdot$GeV. In this domain, the ratio
$p/\varepsilon$, and consequently the $R_e$ ratio, are
approximately independent of the collision energy and equal 
about 1/3 and 0.56,
respectively.
The model assumes that the deconfinement phase--transition is of the
first order. Thus, there is the mixed phase region, corresponding to
the energy interval 30$\div$60 A$\cdot$GeV.
At the end of the mixed phase the $p/\varepsilon$ ratio reaches
minimum (the ``softest point'' of EoS \cite{Hu:95}).
Thus in the transition energy range
the $R_e$ ratio increases
and reaches its maximum, $R_{e}\approx 0.8$, at the end of the transition
domain. Further on, in the pure deconfined phase,
which is represented by an ideal
quark-gluon gas under bag pressure, the $p/\varepsilon$ ratio
increases and approaches
its asymptotic value 1/3 at the highest SPS energy 160 A$\cdot$GeV. 
An estimate of entropy fluctuations can be obtained form the
analysis of multiplicity fluctuations as proposed in
\cite{GaGoMo}.

At the stage of particle freeze-out, the system's entropy is related to
the mean particle multiplicity. We assume that the multiplicity of
negatively 
charged hadrons is proportional to the system entropy, $S\propto
\overline{N}_-$.
Thus the initial entropy fluctuations are transformed into the the
fluctuations of the {\it mean} multiplicity. It is important to
distinguish these dynamical fluctuations of $\overline{N}_-$ formed at the
initial
stage of A+A reaction, from the statistical fluctuations of $N_-$ around
$\overline{N}_-$ at the freeze-out (see Ref.~\cite{GaGoMo} for details).

In Ref.~\cite{GoGaZo} we study within the SMES the energy dependence of
the dynamical strangeness fluctuations caused by the dynamical energy
fluctuations. We define $\overline{N}_s$ as a total number of
strange quark-antiquark pairs created in A+A collision, and consider the
fluctuation
ratio:
\begin{equation}
R_s~=~\frac{(\delta \overline{N}_s)^2/\overline{N}_s^2}{(\delta
E)^2/E^2}~.
\label{Rs}
\end{equation}
When $T\rightarrow \infty$ the system is in the QGP phase. Strange
(anti)quarks can be considered as massless and the bag constant can be
neglected. Then $\varepsilon \propto T^4$ and $n_s\propto T^3$ and
consequently $d\varepsilon /\varepsilon = 4 \cdot dT/T$ and
$dn_s/n_s=3\cdot dT/T$, which result in $R_s=(3/4)^2\cong 0.56$. In the
confined phase, $T<T_C$, the energy density is still approximately given
by $\propto T^4$ due to the dominant contributions of non-strange hadron
constituents. However, the deppendence of the strangeness density on $T$
is
dominated in this case by the exponential factor, $n_s\propto
\exp(-m_S/T)$, as $T<<m_S\cong 500$~MeV. Therefore, at small $T$ one finds
$d\varepsilon /\varepsilon = 4\cdot dT/T$ and $dn_s/n_s = m_S\cdot
dT/T^2$, so that the ratio $R_s=m_S/(4T)$ decreases with $T$. The
strangeness density $n_s$ is small and goes to zero at $T\rightarrow 0$,
but the fluctuation ratio $R_s$ (\ref{Rs}) is large and goes to infinity
at zero temperature limit.

\section{Signals of Deconfinement: Experimental Results}

\begin{figure}[htb]
\mbox{ \includegraphics[width=80mm]{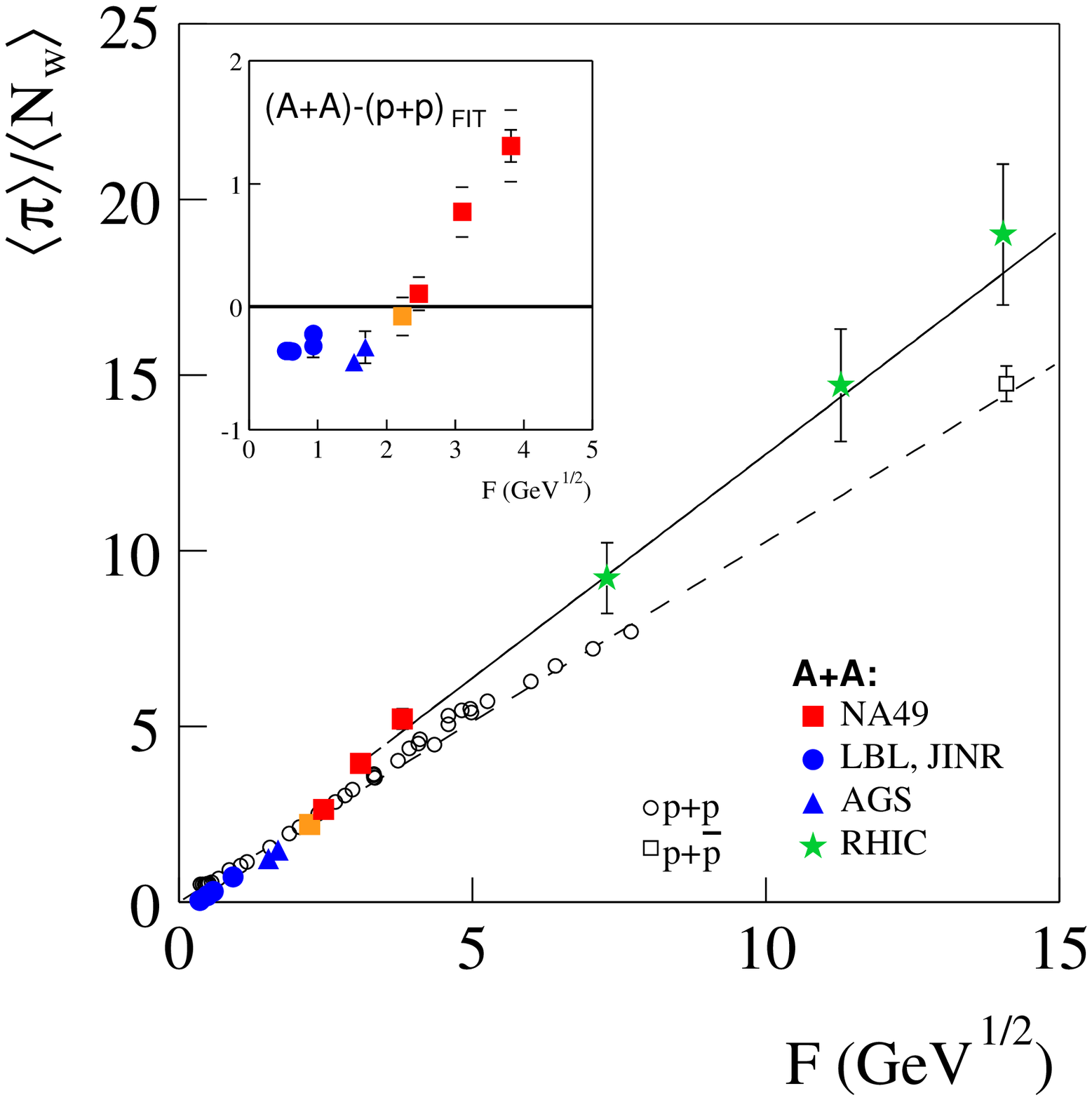} }
\vspace*{-8.0cm}
\narrowcaption[]{.~~The dependence of the total pion multiplicity per
wounded nucleon, $\langle \pi \rangle /\langle N_W\rangle$, on the Fermi's
energy measure $F\equiv(\sqrt{s}-2m_N)^{3/4}\cdot (\sqrt{s})^{-1/4}$
for central A+A collisions (closed symbols) 
and inelastic $p+p(\overline{p})$ interactions
(open symbols). The lines are calculated with SMES \cite{GaGo}.
The dashed line indicates a fit of the form
$a\cdot F$ to the $p+p(\overline{p})$ data ($a=1.01\pm
0.04$~GeV$^{-1/2}$). The solid line
corresponds to the fit of A+A data for initial energy above 40~A$\cdot$GeV 
($a=1.36\pm0.03$~GeV$^{-1/2}$).}
\label{pinp_f}
\end{figure}

\noindent
{\bf The Pion Kink}.~~ A recent compilation of the data on the pion
multiplicity
in central Pb+Pb (Au+Au)
collisions and p+p interactions is shown in Fig.~\ref{pinp_f}.   

\vspace*{-0.3cm}

\vspace*{0.2cm}
\noindent
One observes that the mean pion multiplicity
per wounded nucleon in
$p+p$($\overline{p}$)
interactions is approximately proportional to $F$.
For central A+A collisions, the dependence is  more
complicated and  cannot be fitted
by a single linear function.
Below 40~A$\cdot$GeV the ratio $\langle \pi \rangle/\langle N_W \rangle$
in
A+A collisions is lower than in $p+p$ interactions (pion suppression),
while at higher energies $\langle \pi \rangle/\langle N_W \rangle$ is
larger in A+A collisions than in $p+p$($\overline{p}$) interactions
(pion enhancement). In the region between the AGS and the low SPS
energy, the slope changes from 
$a \cong 1.01$~GeV$^{-1/2}$  
(the dashed line in Fig.~\ref{pinp_f})
to $ a \cong  1.36 $~GeV$^{-1/2}$
(the full line in Fig.~\ref{pinp_f}).

The measured increase of the slope for A+A collisions, by a factor of
about 1.3, is interpreted  within the SMES
as due to an increase of the effective
number of the internal degrees of freedom by a factor of
(1.3)$^4$ $\cong$ 3 and is caused by the creation of a transient
state of deconfined matter at energies higher than 30 A$\cdot$GeV.
%
A transition from the pion suppression \cite{supp} to pion enhancement
is demonstrated more clearly in the insert of Fig.~\ref{pinp_f}, where
the difference between $\langle \pi \rangle/\langle N_W \rangle$
for A+A collisions and the straight line parametrisation of the $p+p$ data
is plotted as a function of $F$
up to the highest SPS energy.

\begin{figure}[htb]
\vspace*{-0.2cm}
\mbox{ \includegraphics[width=60mm]{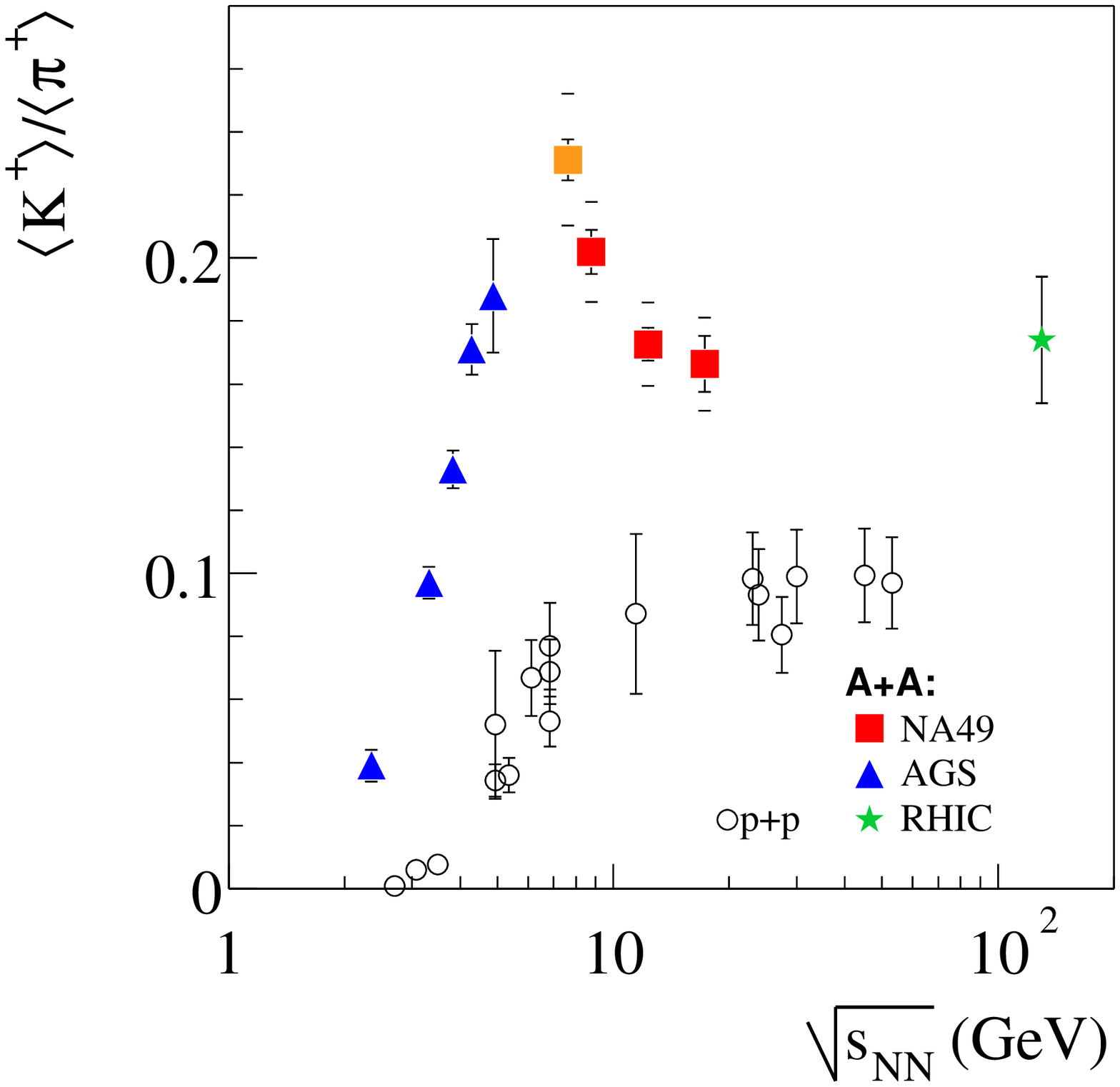}
       \includegraphics[width=60mm]{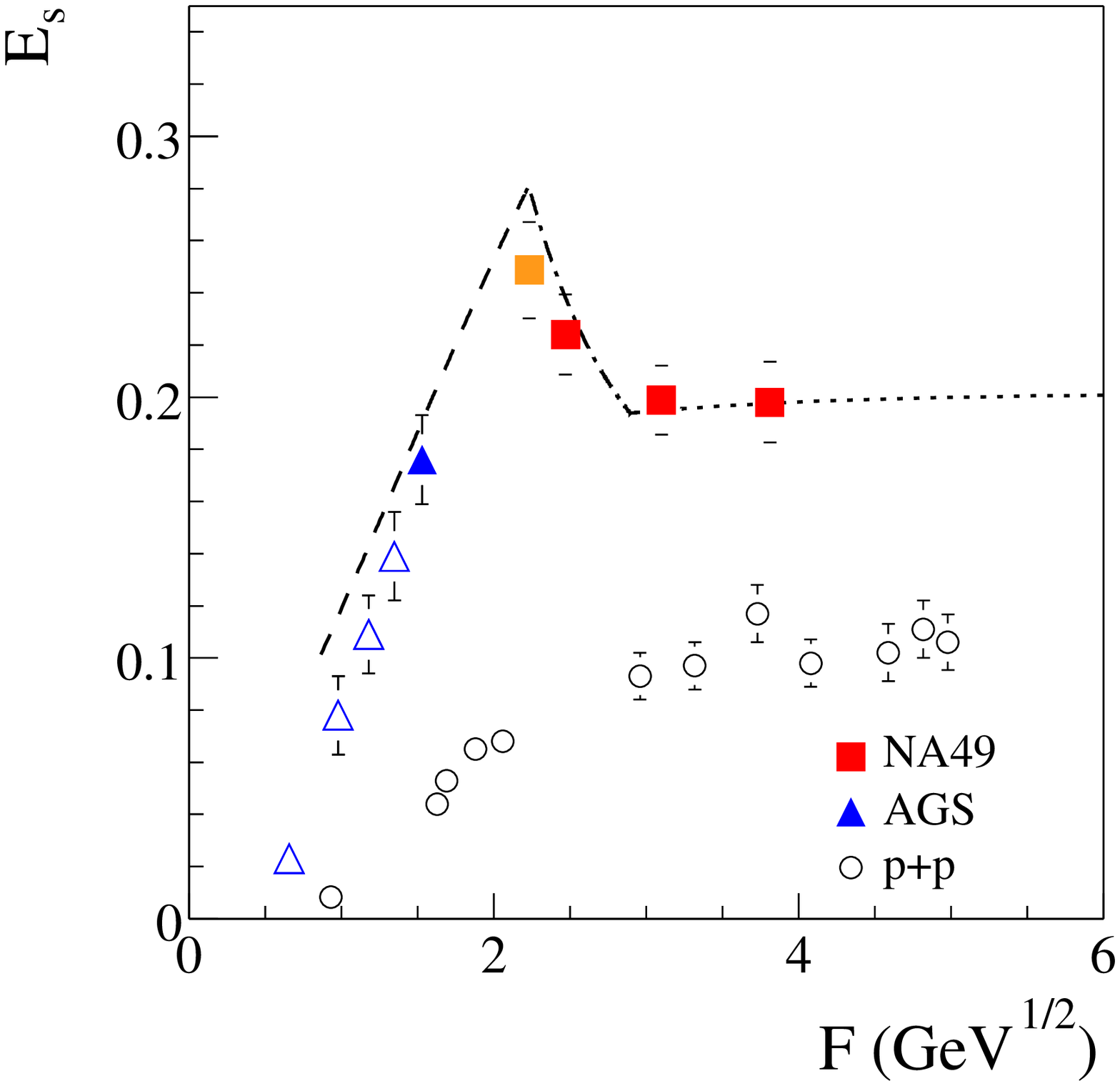} }
%
%
\vspace*{-0.9cm}  
\caption[]{
The dependence of the $\langle K^+ \rangle/\langle \pi^+ \rangle$
(left)
and $E_S$ (right) ratios on the collision energy for
central A+A collisions (closed symbols)  
and inelastic $p+p$ interactions
(open symbols).
The predictions of SMES \cite{GaGo} for the $E_S$ ratio are shown by a line.
Different line styles indicate predictions in the energy
domains in which the confined matter (dashed line), the mixed phase
(dashed--dotted line) and the deconfined matter (dotted line) are created
at
the early stage of the A+A collision.
}
\label{es}
\end{figure}

\noindent
{\bf The Strange Horn}.~~~~ One can argue that the strangeness to entropy
ratio is closely
proportional to the two ratios directly measured in experiments:
the $\langle K^+ \rangle/\langle \pi^+ \rangle$ ratio and
the $E_S = (\langle \Lambda \rangle + \langle K+\overline{K} \rangle)/
\langle \pi \rangle$ ratio.
The energy dependence of both ratios
is plotted in Fig.~\ref{es}
for central Pb+Pb (Au+Au) collisions and p+p interactions.

\vspace*{-0.2cm}

For p+p interactions both ratios show monotonic increase
with energy.
However, very different behavior is observed for central
Pb+Pb (Au+Au) collisions.
The steep threshold rise of the ratio characteristic for confined
matter then settles into saturations at the level expected
for deconfined matter.
In the transition region (at low SPS energies) a sharp maximum
is observed caused by a higher strangeness to entropy ratio
in the confiend matter than in the deconfiend matter.
As seen in Fig.~\ref{es}
the measured dependence is consistent with that expected 
within the SMES.

\begin{figure}[htb]
\vspace*{-0.7cm}
\mbox{ \includegraphics[width=60mm]{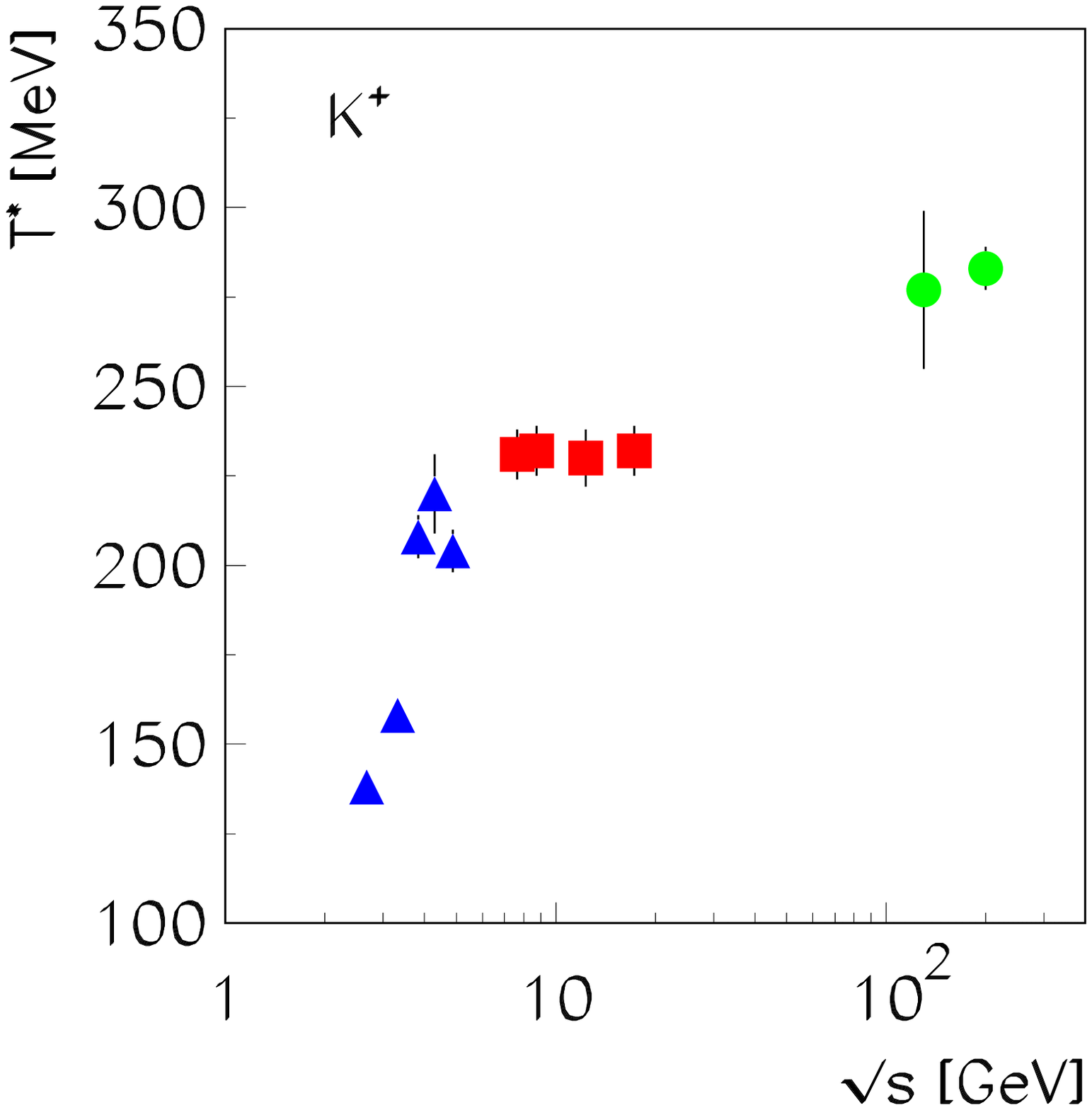}
\includegraphics[width=60mm]{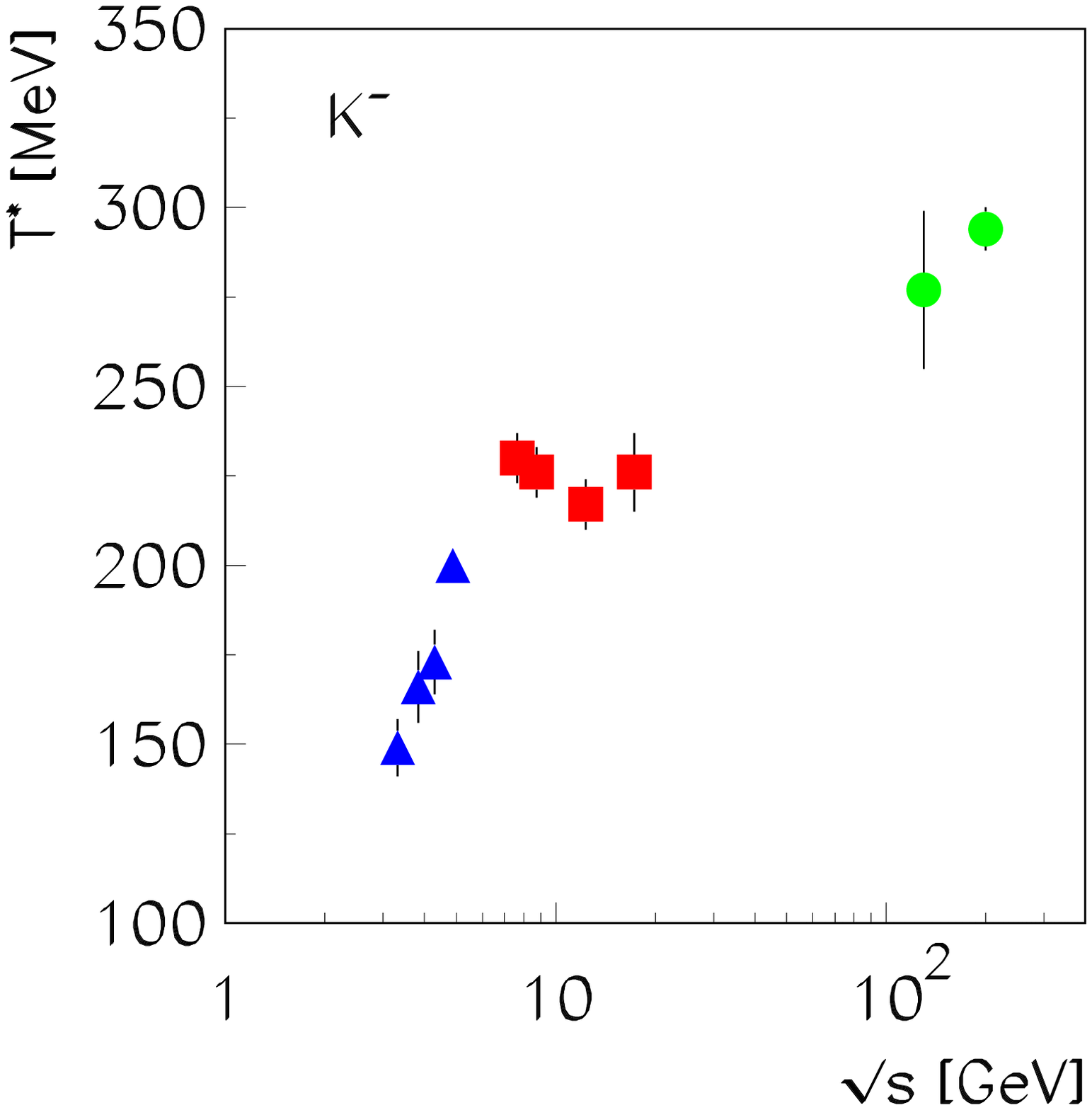} }
\vspace*{-0.5cm}
\caption[]
{The energy dependence of the inverse slope parameter $T^*$ for $K^+$
(left) and $K^-$ (right) mesons (see Ref.~\cite{GoGaBu})  produced at mid-rapidity in
central Pb+Pb (Au+Au) collisions at AGS (triangles), SPS (squares) and
RHIC (circles) energies. }
\label{slopes}
\end{figure}

\vspace{0.2cm}
\noindent
{\bf The Step in Slopes}.~~~~ The energy dependence of the inverse slope
parameter
fitted to  
the $K^+$  and $K^-$  transverse mass
spectra for central Pb+Pb (Au+Au)
collisions is shown in Fig.~\ref{slopes} \cite{GoGaBu}.
The striking features of the data can be summarised and interpreted as
follows.
The $T^{*}$ parameter increases strongly with collision energy up
to the lowest
(30 A$\cdot$GeV) SPS energy point.
This is an energy region where the creation of confined matter at
the early stage of the collisions is expected.
Increasing collision energy leads to an increase of the
early stage  temperature and pressure.
Consequently  the  transverse activity of produced hadrons,
measured by the inverse slope parameter, increases with increasing energy.
The $T^{*}$ parameter is approximately independent
of the collision
energy in the SPS energy range.
In this energy region the transition between confined and deconfined
matter
is expected to be located.
The resulting modification of the  equation of state
``suppresses'' the hydrodynamical transverse expansion and
leads to the observed plateau structure in
the energy dependence of the $T^*$ parameter.
At higher energies (RHIC data),  $T^{*}$ again increases with
the collision
energy. The equation of state  at the early stage  becomes again stiff,
the  early stage temperature and pressure  increase with collision energy,
and this results in increase of $T^{*}$ too.

\vspace*{-0.2cm}

Kaons  
are the best and unique particles among measured hadron species for
observing the effect of the
modification of the equation of state due to the onset
of the deconfinement in hadron transverse momentum spectra.
The arguments are the following.

1). The kaons $m_{T}$--spectra are only weakly affected by the hadron
re-scattering and resonance decays during the post-hydrodynamic hadron
cascade at the SPS and RHIC energies \cite{Sh}. 

2). A simple one parameter exponential fit (\ref{mt}) is quite
accurate for kaons in central A+A collisions at all
energies. This simplifies strongly an analysis of the experimental data.

3). The high quality data on $m_T$-spectra of
$K^+$ and $K^-$ mesons in central Pb+Pb (Au+Au)
collisions are available in the full range of
relevant energies.

\begin{figure}[htb]
\vspace*{-0.5cm}
\mbox{ 
       \includegraphics[width=60mm]{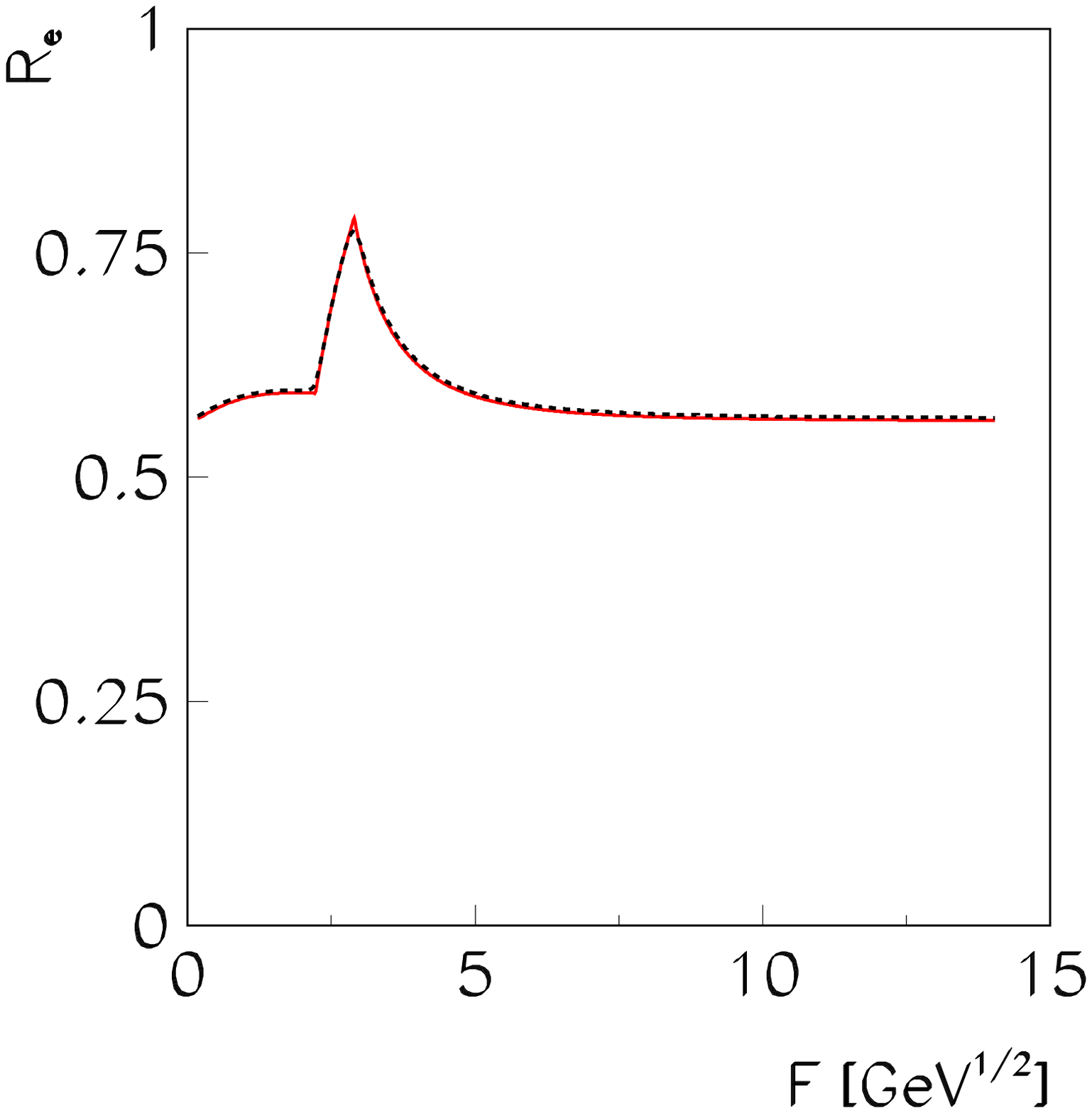}
       \includegraphics[width=60mm]{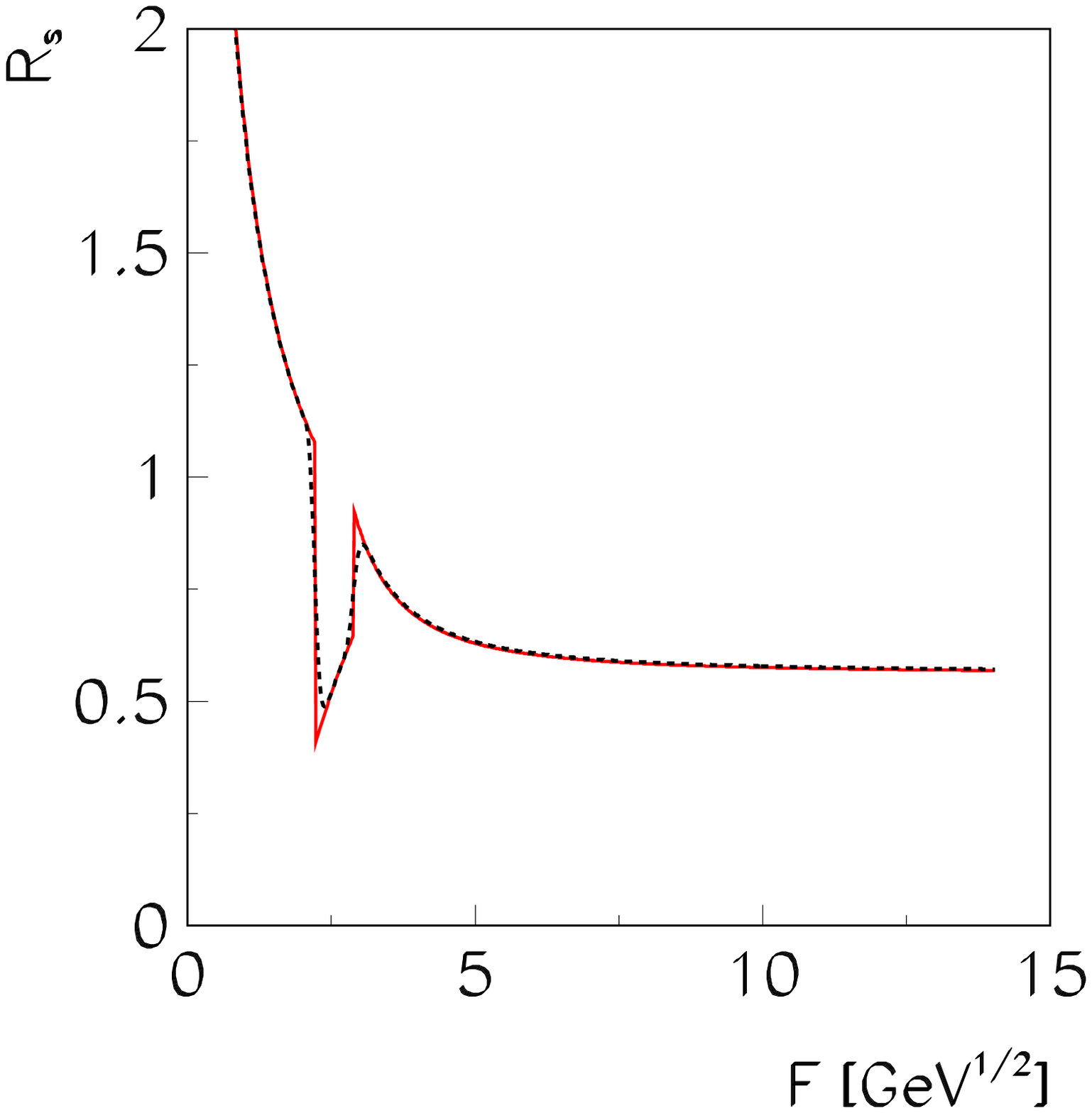} 
}
%
\vspace*{-0.5cm}
\caption[]{
The collision energy dependence of the relative entropy
to energy fluctuations $R_e$ (\ref{R}), left (see Ref.~\cite{GaGoMo}), 
and strangeness to energy 
fluctuations $R_s$ (\ref{Rs}), right (see Ref.~\cite{GoGaZo}),
calculated within SMES.
}
\label{rf}
\end{figure}

\vspace{0.2cm}
\noindent
{\bf The Dynamical Event-by-Event Fluctuations}. ~~~~The
ratios of
entropy to energy (\ref{R}) and strangeness to energy (\ref{Rs})
dynamical fluctuations calculated within SMES  
are presented in Fig.~\ref{rf}.
We find a non-monotonic energy dependence of $R_e$ with a maximum
at the boundary between the mixed phase and the QGP \cite{GaGoMo}. A
pronounced
minimum-structure is expected in the dependence of $R_s$ on 
the collision
energy \cite{GoGaZo}. It is located at $30\div 60$~A$\cdot$GeV, where the
mixed phase is created at the early stage of A+A collision. 


%

The experimental data on the energy dependnece of $R_e$ and $R_s$
ratio are not yet available. Both entropy and strangeness fluctuation
measures, $R_e$ and $R_s$, show anomalous behavior in the transition
region: the maximum is expected for $R_e$ and the minimum for $R_s$.
Consequetly, even a stronger anomaly is predicted for the ratio:
\begin{equation}
R_{s/e}~\equiv~\frac{R_s}{R_e}~=\frac{(\delta
\overline{N}_s)^2/\overline{N}_s^2}{(\delta
\overline{N}_-)^2/\overline{N}_-^2}~.
\label{Rse}
\end{equation}
Experimental measurements of $R_{s/e}$ may be easier than  
the measurements of $R_e$ and $R_s$ because the ratio $R_{s/e}$ requires
measurements of particle multiplicities only, whereas both $R_e$ and
$R_s$ involve also measurements of particle energies.


\section{Conclusions}\label{concl}

The energy scan program at the CERN SPS together with the
measurements at lower (LBL, JINR, SIS, BNL AGS) and
higher (BNL RHIC) energies yielded systematic data
on energy dependence of hadron production in central
Pb+Pb (Au+Au) collisions.
%
%
Predicted  signals of the deconfinement phase transition,
namely anomalies in the energy dependence of hadron production --
the {\it pion kink} \cite{GaGo,Ga}, 
{\it strange horn} \cite{GaGo}
and the {\it step in
slope} of $K$-mesons \cite{GoGaBu}
-- are simultaneously observed 
in the same domain of the
low SPS energies and presented  in the left panel of Fig.~5.

\vspace{-0.2cm}
\begin{figure}[htb]
\mbox{\includegraphics[width=65mm]{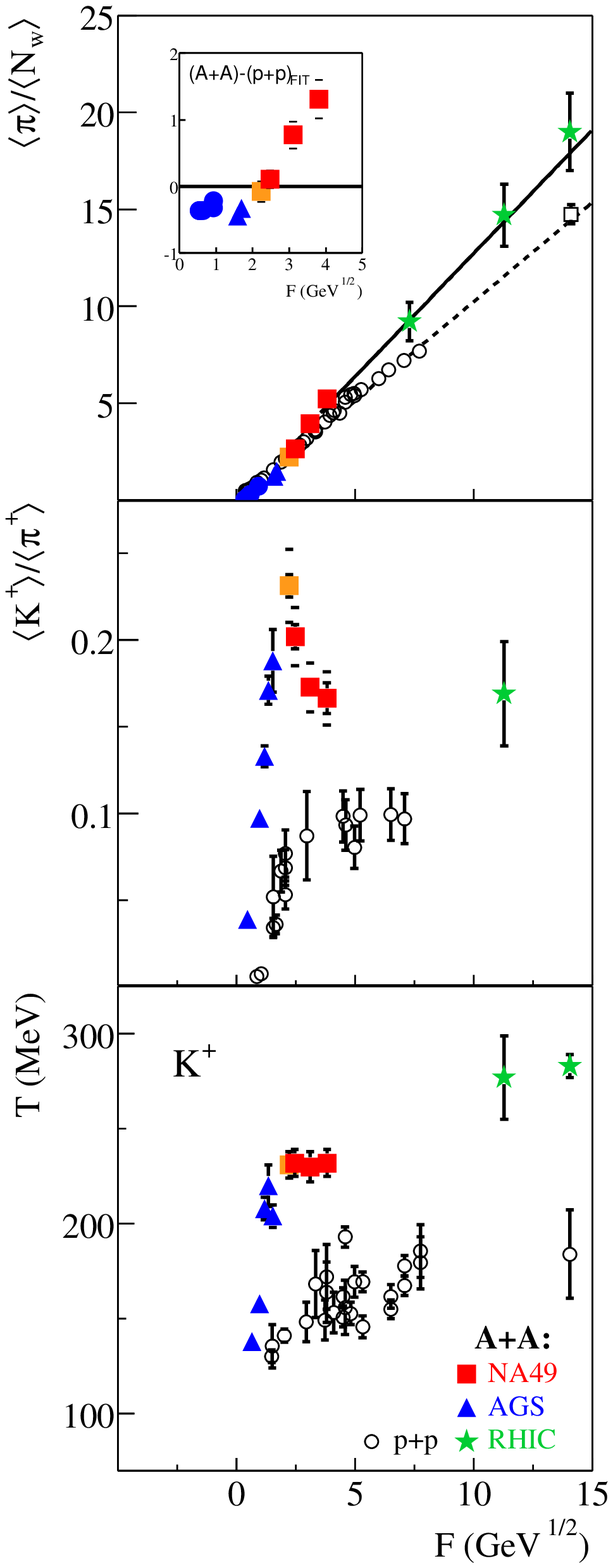}
\includegraphics[width=60mm]{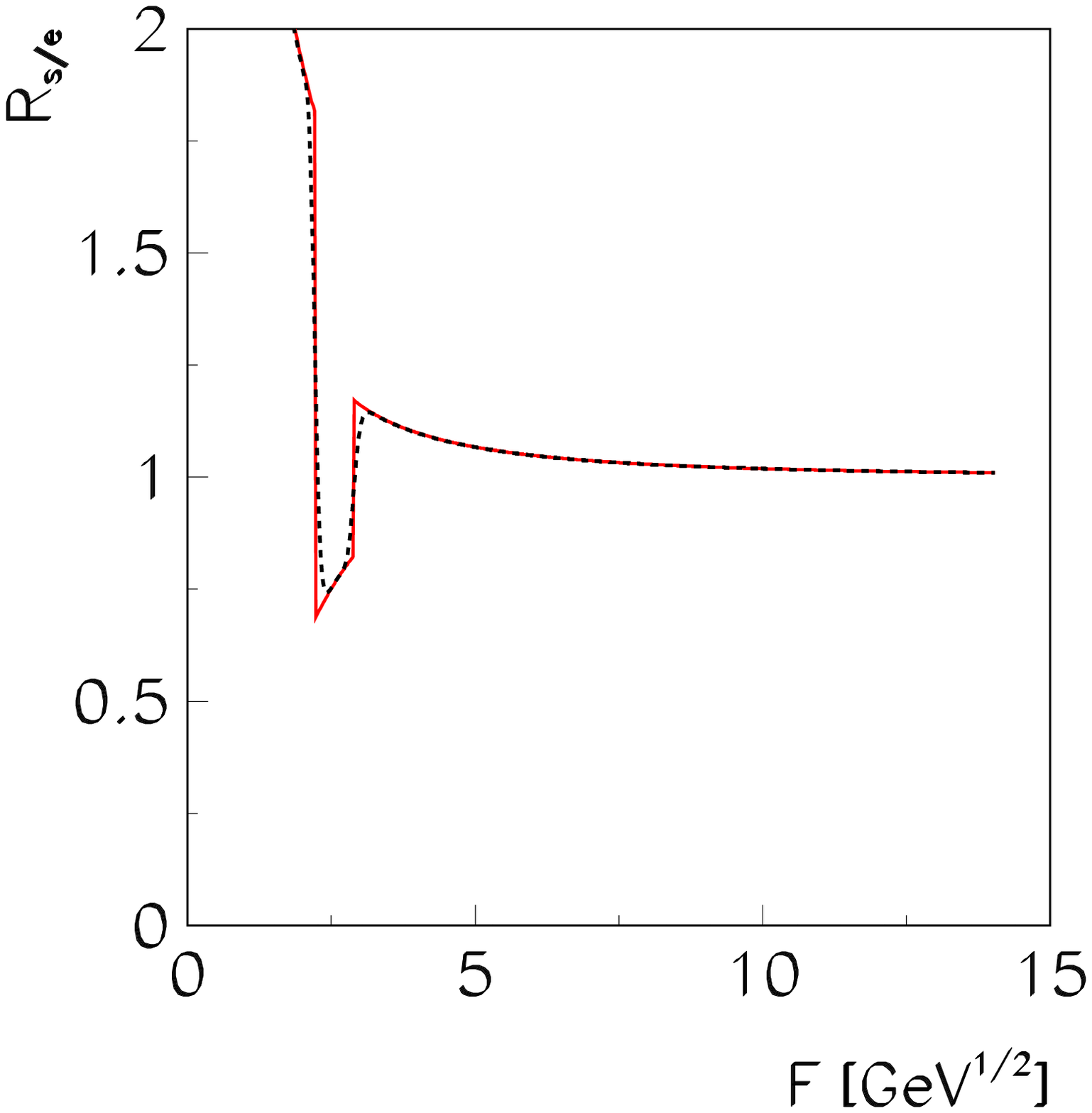}}
\vspace*{-16.0cm}
\narrowcaption[]{.~The collision energy dependence of 
$\langle \pi \rangle/\langle N_W\rangle$, $\langle K^+\rangle/\langle
\pi ^+ \rangle$,
and $T^*(K^+)$. The predicted signature of the deocnfinement transition
-- the pion kink \cite{GaGo,Ga}, the strange horn \cite{GaGo} and the step
in $m_T$-slopes \cite{GoGaBu} are observed simultaneously in
the same domain of collision
energies $F=2\div 3$~GeV$^{1/2}$. The open symbols correspond
to the data in $p+p$ collisions. 
The right panel shows the prediction of
SMES for $R_{s/e}$ (\ref{Rse}) constructed from fluctuations of strange
and non-strange hadron multiplicities \cite{GoGaZo}.
A pronounced minimum of $R_{s/e}$ is predicted in the collision energy
domain in which deconfinement transition is located.} 
 \label{all} \end{figure}

\vspace*{5.5cm}

\noindent
The anomalies in the energy dependence of the hadron production
are seen in central A+A collisions and absent in the data
of $p+p$ reactions. They indicate that the
onset of the deconfinement in central Pb+Pb collisions is located at about
30 A$\cdot$GeV.
The theoretical picture is however far from being complete.
It seems that the Landau-type initial conditions are not appropriated
to describe both the multiplicity and transverse momentum data at the same
time. The initial conditions with spatially spread energy and longitudinal
velocity distributions are needed to reproduce all main features of the
data simultaneously \cite{hama}. The analysis of the data
from the energy scan program is also still in progress. In
particular, first
results at 20~A$\cdot$GeV are soon expected. 
We hope that the properly analysed event--by-event fluctuations
(see Ref.~\cite{GaGoMo})
may also be sensitive to the onset of the deconfinement.
Especially promising looks 
a new measure $R_{s/e}$ (\ref{Rse})
constructed from the fluctuations 
of strange and non-strange hadron multiplicities \cite{GoGaZo}
and shown in the right panel
of Fig.~5.

\vspace{0.5cm}
\noindent
{\bf Acknowledgements.}
The results presented in this report were 
obtained thanks to my close collaboration with Marek
Gaz$^{\! \! \prime}$dzicki.
I would like to thank
him for a fruitful period of joint work. I am also thankful to
Walter Greiner for his permanent support and
encouraging of these studies.

\end{document}